\documentstyle[prl,aps,epsfig]{revtex}
\def\be{\begin{eqnarray}}\def\ba{\begin{eqnarray}}
\def\ee{\end{eqnarray}}\def\ea{\end{eqnarray}}
\def\ben{\begin{enumerate}}\def\bitem{\begin{itemize}}
\def\een{\end{enumerate}}\def\eitem{\end{itemize}}

\def\bi{\bibitem}

\def\prl{Phys. Rev. Lett.}\def\pr{Phys. Rev.}\def\np{Nucl. Phys.}
\def\pl{Phys. Lett.}

\def\Tr{{\mbox{Tr}}}
\def\del{\partial}

\def\roughly#1{\mathrel{\raise.3ex\hbox{$#1$\kern-.75em%
\lower1ex\hbox{$\sim$}}}}
\def\gsim{\roughly>}

\def\la{\langle}\def\ra{\rangle}

\def\A0{A_0}
\def\bq{\begin{equation}}
\def\eq{\end{equation}}

\begin{document}

\twocolumn[%
\hsize\textwidth\columnwidth\hsize\csname@twocolumnfalse\endcsname

\renewcommand{\thefootnote}{\fnsymbol{footnote}}
\setcounter{footnote}{0}

\title{%
 \bf  The Fate of Hadron Masses in Dense Matter: Hidden Local Symmetry
 and Color-Flavor Locking}
\author{{\bf G.E. Brown$^{(a)}$} and {\bf Mannque Rho$^{(b,c)}$}}
\address{
(a) Department of Physics and Astronomy, State University of New
York,  Stony Brook, NY 11794, USA\\ (b) Service de Physique
Th\'eorique, CE Saclay, 91191 Gif-sur-Yvette, France
\\ (c) Physics Departments, Seoul National University and Yonsei University, Seoul, Korea }
 \maketitle
\begin{abstract}
The notion that hadron masses scale according to the scaling of
the quark condensate in hadronic matter, partly supported by a
number of observations in finite nuclei, can be interpreted in
terms of Harada-Yamawaki's ``vector manifestation" (VM) of chiral
symmetry. In this scenario, near chiral restoration, the vector
meson masses drop to zero {\it in the chiral limit} with vanishing
widths. This scenario appears to differ from the standard linear
sigma model scenario. We exploit a link between the VM and
color-flavor locking inferred by us from lattice data on quark
number susceptibility (QNS) measured as a function of temperature
to suggest that local flavor symmetry gets mapped to color gauge
symmetry at the chiral phase transition.
\end{abstract}
\vskip 0.1cm 
 \vskip1pc]
\renewcommand{\thefootnote}{\arabic{footnote}}
\setcounter{footnote}{0}

\section{Evidence from Finite Nuclei}
There are growing indications that as density increases in nuclear
medium, ``hadron masses" drop as proposed in \cite{BR91}. For
instance, we have argued~\cite{BR2001} that the best indication
comes from nuclear response functions such as the longitudinal
response function in $(e.e^\prime p)$ process in
nuclei~\cite{morgenstern}, axial-charge transitions in heavy
nuclei~\cite{BR2001}, nuclear gyromagnetic
ratio~\cite{landau-migdal,schumacher} etc. However these evidences
involve virtual particles and here the ``mass" that appears is a
parameter of an effective field theory and need not be a physical
quantity -- as the pole mass is -- except when radiative
corrections are ignorable. Whether or not radiative corrections
are important depend on the processes considered. The scaling mass
for the $\rho$ meson as discussed in \cite{BR91} has been
invoked~\cite{LKB} to explain the CERES lepton pair data at
invariant masses $\sim 300 - 500$ MeV. Although the explanation
seems viable, it is not unique: there are alternative mechanisms
such as increased widths in medium~\cite{rapp-wambach},
quark-gluon plasma~\cite{alam01} etc. that can equally well
explain the global shift in the invariant mass. It is possible
that a combination of observables in heavy ion processes such as
lepton pair, direct photon... will weed out wrong mechanisms among
the multitude that are present in the literature, see e.g.
\cite{alam01}.

What is clear in the midst of the confusion is that the density
does modify the properties of hadrons. The recent KEK experiment
on the invariant mass spectra of the $e^+e^-$ pairs in the target
rapidity region of 12 GeV $p+A$ reactions does show clearly that
the vector mesons have modified properties in medium~\cite{ozawa}.

In this paper, we exploit the recent developments to describe how
the properties of hadrons may change as the critical density for
chiral phase transition is approached from below. Our arguments
rely on two recent suggestions that come from seemingly unrelated
sectors. One is the suggestion by Harada and Yamawaki~\cite{VM}
that the phase transition from the Nambu-Goldstone phase to the
Wigner-Weyl phase involves ``vector manifestation (VM)" of chiral
symmetry. The other is the proposal by Berges and
Wetterich~\cite{wetterich,berges-wett} that color and flavor get
completely locked even in the Nambu-Goldstone phase. The former
involves low-energy effective field theory and the latter
nonperturbative aspects of QCD {\it proper}.
\section{Hint from Lattice}
It was suggested in \cite{BR96} that the quark number
susceptibility (QNS) $\chi_\pm=(\del/\del\mu_u\pm \del/\del\mu_d)
(\rho_u\pm \rho_d)$ where $\rho_{u,d}$ and $\mu_{u,d}$ are,
respectively, $u,d$-quark number density and chemical potential
measured on lattice as a function of temperature~\cite{lattice}
exhibited a smooth and rapid change-over in {\it both} isoscalar
and isovector channels from a hidden flavor gauge symmetry  to QCD
color gauge symmetry at the chiral transition temperature $T_c$,
implying an intricate connection between the {\it induced}
symmetry and the {\it fundamental} symmetry. When interpreted in
terms of the hidden gauge vector decoupling proposed by
Kunihiro~\cite{hatsuda}, the lattice results on QNS point to a
link between Harada-Yamawaki's ``vector manifestation" in hidden
local symmetry (HLS) (defined precisely below) and color-flavor
locking (CFL) in QCD approaching chiral restoration from the
Goldstone phase and offer a novel interpretation of BR
scaling~\cite{BR91}.
\section{Hidden Local Symmetry in Hadron Sector}
Consider first the hidden local symmetry (HLS) theory of Bando et
al~\cite{bandoetal}. We shall consider -- in the chiral limit
unless otherwise noted -- the symmetry group $[U(2)_L\times
U(2)_R]_{global}\times [U(2)_V]_{local}$ consisting of a triplet
of pions, a triplet of $\rho$-mesons and an
$\omega$-meson~\footnote{The Harada-Yamawaki argument for the VM
scenario is strictly valid for three massless flavors but not
necessarily for two flavors. Just as the phase transition is known
to be different for the two- and three-flavor QCD in both
temperature and density, the two-flavor situation may well be
quite different from the three-flavor scenario. The attitude we
take here is that we are focusing on the non-strange sector of the
three-flavor consideration.}. Motivated by the observation that in
the vacuum ($T=n=0$ where $T$ and $n$ are, respectively,
temperature and density), the $\rho$ and $\omega$ mesons are
nearly degenerate and the quartet symmetry is fairly good
phenomenologically and that near the chiral restoration critical
temperature $T_c$, the $\chi_{\pm}$ behave identically within the
error bars, we put them into a $U(2)$ multiplet. In this theory,
baryons (proton and neutron) do not appear explicitly. They can be
considered as having been integrated out. If needed, they can be
re-introduced in a way consistent with HLS.

The relevant degrees of freedom in the HLS theory are the left and
right chiral fields denoted by $\xi_{L,R}$ and the hidden local
gauge fields denoted by $V_\mu\equiv V_\mu^\alpha
T^\alpha=\frac{\tau^a}{2}\rho_\mu^a +\frac 12 \omega_\mu$ with
$\Tr(T^\alpha T^\beta)=\frac 12\delta^{\alpha\beta}$. If we denote
the $[U(2)_L\times U(2)_R]_{global}\times [U(2)_V]_{local}$
unitary transformations by $(g_L,g_R,h)$, then the fields
transform $\xi_{L,R}\mapsto h(x)\xi_{L,R}g^\dagger_{\small{L,R}}$
and $V_\mu\mapsto h(x)(V_\mu-i\del_\mu)h^\dagger (x)$. This theory
has the correct symmetry structure as well as dynamical contents
for low-energy excitations of hadrons. Symmetry considerations
alone however do not give a unique phase structure of the theory.
In fact with any given parameters of the theory, it has a
multitude of flow structure as one decimates down in the Wilsonian
sense~\cite{HYfate}. It is however by matching with QCD at the
chiral scale $\Lambda_\chi$ that the parameters of the theory and
the phase structure become unique. It was shown in \cite{HYfate}
that among the multitude of the possibilities, it is the vector
manifestation with a consequent strong violation of vector
dominance that is uniquely picked for the chiral phase transition.

We follow \cite{HY:letter,HY:matching} and consider the HLS
Lagrangian as an effective Lagrangian that results when
high-energy degrees of freedom above the chiral scale $\Lambda$
are integrated out. Now the scale $\Lambda$ will in general depend
on the number of flavors $N_f$, density $n$ or temperature $T$
depending upon what system is being considered. This is a {\it
bare} Lagrangian in the Wilsonian sense with the parameters  $g_V
(\Lambda)$ which is the hidden gauge coupling constant,
$a(\Lambda)$ which signals that chiral $U(2)\times U(2)$ symmetry
is spontaneously broken by taking the value $a\neq 1$ and $f_\pi
(\Lambda)$ which is the pion decay constant playing the role of
the order parameter for chiral symmetry with $f_\pi=0$ signaling
the onset of the Wigner-Weyl phase.  These parameters can be
determined~\cite{HY:matching} in terms of QCD condensates by
matching -- \`a la Wilson -- the vector and axial-vector
correlators with the ones of QCD at the chiral scale
$\Lambda_\chi$. By following renormalization group (RG) flows to
low-energy scales, one can obtain low-energy parameters that can
be related to those that figure in chiral perturbation theory. An
important observation here is that the (assumed) equality of the
vector and axial-vector correlators ($\Pi_V=\Pi_A$) at chiral
restoration where $\la \bar{q}q\ra=0$ implies that the HLS theory
approaches the Georgi vector limit~\cite{georgi}, namely, $g_V=0$
and $a=1$, plus the vanishing of $f_\pi$, which is referred to as
``vector manifestation."

The above argument is quite general and should be applicable
equally to temperature, density and $N_f$. In terms of baryon
density $n$, this implies that at the critical density $n=n_c$, we
must have
 \ba
g_V(\Lambda(n_c); n_c)= 0, \ \ f_\pi (\Lambda(n_c); n_c)= 0, \ \
a(\Lambda(n_c); n_c)= 1
 \ea
where we have indicated the density dependence of the cutoff
$\Lambda$.

In HLS theory, the vector masses are given by the Higgs mechanism.
In free space, it is of the form
 \be
m_V\equiv m_\rho=m_\omega =\sqrt{a(m_V)} g_V (m_V) f_\pi
(m_V)\label{ksrf}
 \ee
where the cutoff dependence is understood. Here the parameter
$a(m_V)$ etc means that it is the value at the scale $m_V$
determined by an RG flow from the {\it bare} quantity
$a(\Lambda)$.
Note that (\ref{ksrf}) is similar, but not identical, to the KSRF
relation $m_\rho=\sqrt{2}g_{\rho\pi\pi} f_\pi (0)$. Now in medium
with $n\neq 0$, if we assume that the $U(2)$ symmetry continues to
hold also in density as in temperature, we expect this mass
formula to remain the same except that it will depend upon
density,
 \be
m^\star_V\equiv m^\star_\rho=  m^\star_\omega =
\sqrt{a(m^\star_V)} g_V (m^\star_V) f_\pi
(m^\star_V)\label{ksrfn}.
 \ee
The density dependence is indicated by the star. As in the case of
$N_f$ discussed in \cite{VM}, the cutoff $\Lambda$ will depend
upon density, say, $\Lambda^\star$ understood in (\ref{ksrfn}).

The Harada-Yamawaki argument (or ``theorem") would imply that at
$n=n_c$ where $\la\bar{q}q\ra=0$ and hence $\Pi^\star_V
(n_c)=\Pi^\star_A (n_c)$,  the Georgi's vector limit $g_V=0$,
$a=1$ is reached together with $f_\pi=0$. This means that
 \ba
m_V^\star (n_c)= 0.
 \ea
At this point, the quartet scalars will be ``de-Higgsed" from the
vector mesons and form a degenerate multiplet with the triplet of
massless pions with the massless vectors decoupled. This assures
that the vector correlator is equal to the axial-vector correlator
in the HLS sector matching with the QCD sector, i.e., the ``vector
manifestation" of chiral symmetry. In this scenario, {\it dictated
by the renormalization group equations, the vector meson masses
drop as density increases}. This is in agreement with Adami and
Brown who arrived at the same conclusion in temperature using QCD
sum rules~\cite{adami-brown}. Note that this scenario is distinct
from the ``standard" (as yet to-be-established) picture in which
the $\rho$ and $a_1$ come together as do the pions and a scalar
$\sigma$. In the standard scenario, there is nothing which forces
the vector mesons to become massless and decouple. They can even
become more massive, even in the chiral limit, at chiral
restoration than in the vacuum provided the vector dominance is
assumed ~\cite{pisarski}. Thus the vanishing of the vector-meson
mass together with the breakdown of vector dominance~\cite{HYfate}
is a {\it prima facie} signal for the phase transition \`a la
Harada and Yamawaki.

Harada and Yamawaki showed that such a chiral restoration in the
vector manifestation can occur at some large number of flavors
$N_f^c \gsim 4$, as expected in large $N_f$ QCD. This means $T_c
(N_f^c)=0$. Since $T_c (N_f=0) > N_c (N_f=2) > N_c (N_f=3) ... >
T_c (N_f=N_f^c\sim 5)=0$ as observed in lattice gauge
calculations~\cite{karsch}, we infer that the temperature-driven
chiral transition also could involve the VM even though the
sigma-model scenario is currently in vogue at least for $N_f=2$
among the workers in the field~\cite{karsch}.
\section{Color Flavor Locking}
Quite remarkably the scenario given above in terms of effective
field theory of strong interactions can be mapped to QCD
condensates. It has been recently shown~\cite{raja-wilcz} that at
asymptotically high density, the color $SU(3)_c$ and the flavor
$SU(3)_F$ get locked completely through diquark condensates
leading to color-flavor-locked (CFL) superconductivity. What is
even more surprising is that the CFL phenomenon can take place
also in the hadronic phase (at low density).

Now how the CFL would manifest in the vacuum phase depends on
whether the strange-quark mass is considered as ``heavy" or
``light" compared with the non-strange chiral
quark~\cite{wetterich}. Symmetry considerations alone would imply
completely different pictures implying a phase transition between
the two at some strange quark mass $m_s^c$. Dynamics must however
make the real situation lie somewhere in between in a way quite
analogous to the skyrmion description of the hyperons. Keeping
this point in mind, let us consider the two-flavor case. In
\cite{berges-wett}, Berges and Wetterich argue that both the
quark-antiquark condensate $\chi$ defined by
 \be \chi=\la\bar{q}_\alpha^a\sum_{i=1}^3
(\tau_i)_{\alpha\beta} (\lambda_i)^{ab} q_\beta^b\ra\label{chi}
 \ee
and the diquark condensate $\Delta$ defined by
 \be
\Delta=\la q_\alpha^a (\tau_2)_{\alpha\beta} (\lambda_2)^{ab}
q_\beta^b\ra\label{delta}
 \ee
(where the indices $\alpha, \beta$ denote the flavors and $a, b$
the colors) can be nonzero in the Goldstone mode and suggest that
one or both of them ``melt'' at chiral restoration in conjunction
with deconfinement. Color is completely broken by the two
condensates (\ref{chi}) and (\ref{delta}). This pattern of color
breaking and color-flavor locking renders the octet gluons and six
quarks massive by the Higgs mechanism and generates three
Goldstone pions due to the broken chiral symmetry. All of the
excitations are integer-charged. Among the eight massive gluons,
three of them are identified with the isotriplet $\rho$'s with the
mass
 \be
m_\rho=\kappa g_c \chi\label{cfl-rho}
 \ee
where $\kappa$ is an unknown constant and $g_c$ the color gauge
coupling. The fourth vector meson is identified with the
isosinglet $\omega$ with the mass
 \be
m_\omega=\kappa^\prime g_c \Delta\label{cfl-omega}
 \ee
where $\kappa^\prime$ is another constant. The remaining four
vector mesons which must be ``heavy" can be given a physical
meaning only when the strangeness quantum number is taken into
account~\cite{wetterich}. As for the fermions, there are two
baryons with the quantum numbers of the proton and neutron with
their masses proportional to the scalar condensate $\phi$, $
\phi=\la \bar{q}_\alpha^a q_\alpha^a\ra$. The four remaining
fermions must also be ``heavy" and can be interpreted as
(physical) hyperons when strangeness is implemented as in the case
of the vector mesons. What concerns us in this note is therefore
the three pions, the proton and neutron, the $\rho$-mesons and the
$\omega$-meson.

As in \cite{BR96,hatsuda}, we interpret the ``measured" singlet
and non-singlet QNS's~\cite{lattice} to indicate that both the
$\rho$ and $\omega$ couplings vanish or nearly vanish at the
transition temperature $T_c$. This means that the $\omega NN$
coupling which is $\sim 3$ times the $\rho NN$ coupling at zero
temperature become equal to the latter at near the critical
temperature. This also means that the both vector mesons become
massless  and decoupled. Viewed from the CFL point of view, it
follows from (\ref{cfl-rho}) and (\ref{cfl-omega}) that the
condensates $\chi$ and $\Delta$ ``melt" at that point. This is
consistent with the observation by Wetterich~\cite{wetterich} that
for three-flavor QCD, the phase transition -- which is both chiral
and deconfining -- occurs at $T_c$ with the melting of the
color-octet condensate $\chi$. The transition is first-order for
$N_f=3$ in agreement with lattice calculations, so the vector
meson mass does not go to zero smoothly but makes a jump from a
finite value to zero. We expect however that in the case of
$N_f=2$ the transition will be second-order with the vector mass
dropping to zero continuously.

Combined with the HLS and CFL results ((\ref{cfl-rho}),
(\ref{cfl-omega}) and (\ref{ksrf}) -- all of which are Higgsed),
the lattice results invite us to set
 \be
ag_V f_\pi\approx \kappa g_c\chi \approx \kappa^\prime
g_c\Delta.\label{crossover} \ee
 Since in the case of $SU(3)_f$, the diquark condensate $\Delta$ must
be zero, while the equality of the $\rho$ and $\omega$ masses is
to still hold, this means that in the above expression, $\Delta$
must be replaced by $\chi$ in going from $SU(2)_f$ to $SU(3)_f$.
This change may be interpreted as a phase
transition~\cite{wetterich}. The main observation of our note is
that the vanishing of the hidden gauge coupling $g_V$ matches onto
the vanishing of the {\it condensates} $\chi$ and $\Delta$ with
the color gauge coupling $g_c\neq 0$.

Next, as shown in \cite{BR96},  above the chiral transition
temperature $T\gsim T_c$, the QNS's can be well described by
perturbative gluon exchange with a small gluon coupling constant
$\frac{g_c^2}{4\pi}\ll 1$. This implies that the flavor gauge
symmetry {\it cedes} to the fundamental QCD gauge symmetry at the
phase transition. We propose that Eq. (\ref{crossover}) describes
the {\it relay} that takes place in terms of the hidden flavor
gauge coupling $g_V$ on one side and the color gauge coupling
$g_c$ on the other side. Now above $T_c$, the color and flavor
must unlock, with the gluons becoming massless and releasing the
scalar Goldstones. The way the two condensates melt as temperature
is increased is a dynamical issue that cannot be addressed within
the present scheme.

At present, unlike hot matter, there is no guidance from lattice
on dense matter. We shall therefore assume that the above scenario
holds in density up to $n=n_c$. As suggested by Sch\"afer and
Wilczek~\cite{sch-wilcz}, one possibility is that the three-flavor
color-flavor locking operative at asymptotic density continues all
the way down to the ``chiral transition density" ($n_c$) in which
case there will be no real phase change since there will then be a
one-to-one mapping -- although symmetries are ``twisted" --
between hadrons and quark/gluons, e.g., in the sense of
hadron-quark continuity. In nature however the non-negligible
strange-quark mass is likely to spoil the ideal three-flavor
consideration. The alternative scenario that we adopt here is that
viewed from ``bottom-up," the hadronic phase with $\chi\neq 0$,
$\sigma\neq 0$ and $\Delta\neq 0$ goes over to that with
$\chi=\sigma=0$ and $\Delta\neq 0$ corresponding to the two-flavor
color superconducting (2csc) phase~\cite{raja-wilcz}. In this case
we will preserve the mass formula (\ref{ksrf}) as one approaches
$n_c$.
\section{BR scaling and Landau Fermi Liquid}
In the Harada-Yamawaki scenario, at density approaching critical,
the width should become narrower, decreasing like $\sim g_V^2$,
with increasing density. Then the vector mesons would behave more
like a quasiparticle at higher density than at lower density. {\it
This is our new interpretation of the BR scaling originally
formulated with the Skyrme Lagrangian}. The situation near the
normal matter density is undoubtedly more complicated.
Nonetheless, several cases evidencing BR scaling are discussed in
a recent review~\cite{BR2001}. Some are somewhat model-dependent
and hence subject to objections. The most direct case is the
$(e,e^\prime p)$ response functions in
nuclei~\cite{morgenstern,BR2001} where the effect of BR scaling is
more prominently exposed.

Thus far, we have argued for a link between the
color-flavor-locked condensates and hidden gauge symmetry. This
will constitute a major progress in the field if confirmed. Here
we propose an even more remarkable connection between QCD
``vacuum" properties encoded in BR scaling and many-body nuclear
interactions. This comes about because nuclear matter owes its
stability to a Fermi-liquid fixed point~\cite{landau-migdal} as a
consequence of which certain interesting nuclear properties turn
out be calculable in terms of the Fermi-liquid fixed point
parameters. Specifically, it has been shown that the Landau
parameter $F_1$ -- which is a component of quasiparticle
interactions -- can be expressed in terms of the BR scaling factor
$\Phi (n)\equiv m_\rho^\star (n)/m_\rho (0)$. An observable that
probes this relation is the anomalous gyromagnetic ratio $\delta
g_l$ in heavy nuclei which takes the form~\cite{landau-migdal}
 \be
\delta g_l=\frac{4}{9}\left[\Phi^{-1}-1-\frac 12\tilde{F}_1
(\pi)\right]\tau_3\label{deltagl}
 \ee
where $\tilde{F}_1 (\pi)$ is the pionic contribution to $F_1$
which is completely given for any density by chiral symmetry. We
should stress that (\ref{deltagl}) is valid only near nuclear
matter density. At nuclear matter density, it takes the value
$\tilde{F}_1 (\pi)|_{n=n_0}=-0.153$. Note that (\ref{deltagl})
depends on only one parameter, $\Phi$. This parameter can be
extracted either from nuclear matter saturation ($m_\omega^\star$)
or from Gell-Mann-Oakes-Renner formula for in-medium pion
($f_\pi^\star$) or from a QCD sum rule for the $\rho$ meson
($m_\rho^\star$). All give about the same value, $\Phi
(n_0)\approx 0.78$. Given $\Phi$ at nuclear matter density,
Eq.(\ref{deltagl}) makes a simple and clear-cut prediction,
$\delta g_l=0.23\tau_3$. This was confirmed by a measurement for
proton in the Pb region~\cite{schumacher}, $\delta g_l^p=0.23\pm
0.03$. Turning the reasoning around, we could consider this a
quantitative determination of the scaling factor $\Phi$ at nuclear
matter density.

We are grateful for discussions with, and comments from, Masayasu
Harada and Koichi Yamawaki. The work of GEB was supported by the
US Department of Energy under Grant No. DE-FG02-88 ER40388 and
that of MR in part by the Brain Korea 21 in 2001.


\end{document}